\documentstyle[prd,aps]{revtex}
\begin{document}
\draft
\twocolumn[\hsize\textwidth\columnwidth\hsize\csname
@twocolumnfalse\endcsname
\preprint{SU-ITP-95-21, UH-IfA-95/51, hep-th/9510119}
\title{  Nonthermal   Phase Transitions After
Inflation}
\author{Lev Kofman$^1$, Andrei Linde$^2$, and Alexei A. Starobinsky$^3$}
\address{$^1$Institute for Astronomy, University of Hawaii,
2680 Woodlawn Dr., Honolulu, HI 96822, USA}
\address{$^2$Department of Physics, Stanford University, Stanford, CA
94305, USA}
\address{$^3$Landau Institute for Theoretical Physics,
Kosygina St. 2, Moscow 117334, Russia}
\date{October 17, 1995}
\maketitle
\begin{abstract}
  At the first stage of reheating after inflation, parametric resonance may
rapidly transfer most of the energy of an
inflaton field $\phi$  to the energy of other  bosons.  We show that quantum
fluctuations of  scalar and vector fields
produced at this stage  are much greater than they  would be
in a state of thermal equilibrium. This   leads to   cosmological
phase transitions of a new type, which may result in a copious
production of topological defects and in  a
secondary stage of  inflation after  reheating.
\end{abstract}
\pacs{PACS: 98.80.Cq, 04.62.+v, 05.70.Fh \hskip 1.8cm SU-ITP-95-21,
UH-IfA-95/51,
{}~hep-th/9510119}
\vskip2pc]

 The theory of reheating is one of the most important parts of
inflationary
cosmology. Elementary theory of this process was developed many years
ago
\cite{DL,st81}.   Some important steps toward a complete theory have
been
made in \cite{Brand}.    However,  the real progress in
understanding of
this process was achieved only recently when the new theory of
reheating was
developed.  According to this theory   \cite{KLS}, reheating
typically consists
of  three different stages.  At the first stage, a classical
 oscillating scalar  field $\phi$ (the  inflaton field) decays into
massive
bosons due to parametric
resonance.  In
many models the resonance is very broad, and the process occurs
extremely
rapidly.  To distinguish this stage of explosive reheating from the
stage of  particle decay and
thermalization, we called it {\it preheating}.
Bosons produced at that stage are far away from thermal equilibrium
and have enormously large occupation numbers. The second stage
is the
decay of previously produced particles. This stage typically can be
described
by methods developed in   \cite{DL}. However, these methods should be
applied
not  to the decay of the original homogeneous inflaton field, but to
the decay
of particles produced at the  stage of preheating. This
 changes many features of the process including the final value of
the
reheating temperature.  The third stage of reheating is
thermalization.

Different aspects of the theory of explosive reheating have been
studied  by many  authors  \cite{Shtanov} --\cite{Kaiser}.
In our presentation we will follow  the original approach of ref.  \cite{KLS},
where the theory of reheating was investigated with an account taken  both  of
the expansion of the universe and of the backreaction of created particles.

One should note that
there exist such models where this first stage of reheating is
absent; e.g, there is no parametric resonance in the theories where
the
 field $\phi$ decays into fermions. However, in the theories where
preheating is possible one may expect  many  unusual phenomena.
One of the most
interesting  effects  is the
possibility of specific nonthermal post-inflationary phase
transitions which occur after preheating. As we
will see, these phase transitions in certain cases can be much more
pronounced
that the standard high temperature cosmological phase transitions
\cite{Kirzhnits,MyBook}. They may lead to copious
production of topological defects and to a
secondary stage of  inflation after  reheating.

 Let us first remember the  theory of  phase transitions in
theories
with spontaneous symmetry breaking in the theory of scalar
fields
$\phi$ and $\chi$ with the effective potential
\begin{equation}\label{p1}
V(\phi,\chi) =   {\lambda\over 4}
(\phi^2-\phi_0^2)^2 + {
1\over2} g^2 \phi^2 \chi^2   \ .
\end{equation}
  Here $\lambda, g \ll 1$  are
coupling constants. $V(\phi,\chi)$ has a minimum at $\phi =
\phi_0$,
$\chi = 0$ and a maximum at $\phi = \chi = 0$ with the curvature
$V_{\phi\phi}
= -m^2= - \lambda\phi_0^2$. This effective potential
acquires corrections due  to quantum (or
thermal)
fluctuations of the scalar fields \cite{Kirzhnits,MyBook},
$
\Delta V =  {3\over 2} \lambda  \langle (\delta\phi^2)\rangle \phi^2 +
{g^2\over 2}  \langle (\delta\chi)^2\rangle  \phi^2 + {g^2\over
2}\langle (\delta\phi)^2\rangle
\chi^2 +...,
$
where  the quantum field operator is decomposed as $\hat \phi =
\phi + \delta \phi$ with $\phi \equiv \langle \hat \phi\rangle$,
and we have written only
leading terms depending on $\phi$ and  $\chi \equiv \langle \hat \chi
\rangle$. In
the large temperature limit
$
\langle (\delta\phi)^2\rangle = \langle (\delta\chi)^2\rangle = {T^2\over 12}.
$
The effective mass squared of the field $\phi$
\begin{equation}\label{p4}
m_{\phi ,eff}^2 = -m^2 + 3 \lambda \phi^2 +
3\lambda \langle
(\delta\phi)^2\rangle +  g^2\langle(\delta\chi)^2\rangle
\end{equation}
becomes positive and symmetry  is restored  (i.e. $\phi =0$ becomes
the stable equilibrium point) for $T > T_c$,
where
$T^2_c = {12 m^2\over 3\lambda + g^2} \gg m^2$. At this temperature
the
energy density of the gas of ultrarelativistic particles
is given by
$
\rho = N(T_c) {\pi^2\over 30} T_c^4 = {24\, m^4N(T_c)\pi^2\over 5\,
(3\lambda +
g^2)^2} \
{}.
$
Here $N(T)$ is the effective number of degrees of freedom at large
temperature,
which in realistic situations may vary from $10^2$ to $10^3$.
Note that for   $g^4 < {96 N\pi^2\over 5}
\lambda$ this energy
density   is
greater than the vacuum energy density $V(0) = {m^4\over
4\lambda}$.  Meanwhile, for   $g^4 {\
\lower-1.2pt\vbox{\hbox{\rlap{$>$}\lower5pt\vbox{\hbox{$\sim$}}}}\ }
\lambda$
radiative corrections are important, they lead to creation of a
local
minimum of $V(\phi,\chi)$, and the phase transition occurs from a
strongly
supercooled state \cite{Kirzhnits}. That is why the first models of
inflation
required supercooling at the moment of the phase transition.

An exception from this rule is given by supersymmetric
theories, where
one may have $g^4 \gg \lambda$ and still have a  potential which is
flat near
the origin due to cancellation of quantum corrections of bosons and
fermions
\cite{LS}. In such cases thermal energy becomes smaller than the
vacuum energy
at $T < T_0$, where $T^4_0 = {15 \over 2 N \pi^2}m^2\phi_0^2$. Then
one may
even
have a short stage of inflation which begins at $T \sim T_0$ and ends
at $T =
T_c$. During this time the universe may inflate by the factor
\begin{equation}\label{p5a}
{a_c\over a_0} = {T_0\over T_c} \sim 10^{-1} \Bigr({g^4\over
\lambda}\Bigl)^{1/4}\,
\approx 10^{-1} g\,\sqrt{\phi_0\over  m} .
\end{equation}

In supersymmetric theories with flat
directions $\Phi$
it may be more natural  to consider potentials of the so-called  ``flaton''
fields $\Phi$
without the term  ${\lambda \over 4} \Phi^4$  \cite{LS}:
\begin{equation}\label{p5b}
V(\Phi,\chi) =  - {m^2 \Phi^2\over2}
+{\lambda_1 \Phi^6\over 6 M_{\rm P}^2} + {m^2 \Phi_0^2\over 3} + {1
\over2} g^2 \Phi^2 \chi^2
 \ ,
\end{equation}
where $\Phi_0 =  \lambda_1^{-1/4} \sqrt{m M_{\rm P}}$ corresponds to the
minimum of
this potential.  The critical temperature in this theory for
$\lambda_1 \Phi_0^2
\ll g^2 M_{\rm P}^2$ is   the same   as in the theory (\ref{p1}) for
$\lambda \ll
g^2$, and expansion of the universe during thermal inflation is
given by $10^{-1}
g\,\sqrt{\Phi_0/m}$, as in eq. (\ref{p5a}). This short additional
stage of
``thermal inflation'' may
be very
useful; in particular, it may provide a solution to the Polonyi field
problem \cite{LS}.

The theory of   cosmological phase transitions is an important part
of the
theory of the evolution of the universe, and during the last twenty
years it was investigated in a very detailed way. However, typically
it
was
assumed that the phase transitions occur in the state of thermal
equilibrium.
Now  we are going to show that similar phase transitions
may occur
even much more efficiently  prior to  thermalization, due to the
anomalously
large expectation values $\langle(\delta\phi)^2\rangle$ and
$\langle(\delta\chi)^2\rangle$
produced during preheating.

We will first consider the model (\ref{p1})   without the scalar
field
$\chi$
and with the amplitude of spontaneous symmetry breaking  $\phi_0 \ll
M_{\rm P}$. In this model inflation occurs during the slow rolling of the
scalar
field $\phi$ from its very large values until it becomes of the order
$M_{\rm P}$.
Then it   oscillates with the  initial amplitude $\phi \sim 10^{-1} M_p$ and
initial frequency $\sim 10^{-1} \sqrt  \lambda
M_{\rm P}$.
Within a few dozen  oscillations it transfers most of its energy $\sim
{\lambda\over 4}10^{-4}
M_{\rm P}^4$ to
its long-wave fluctuations $\langle(\delta \phi)^2\rangle$ in the
regime
of broad parametric resonance  \cite{KLS}.

The crucial observation   is the following. Suppose that the initial energy
density of oscillations $\sim {\lambda\over 4}10^{-4}
 M_{\rm P}^4$ were instantaneously transferred to thermal energy density $\sim
10^2  T^4$. This would give the reheating
temperature  $T_r \sim 2\times 10^{-2} \lambda^{1/4} M_{\rm P}$, and the scalar
field  fluctuations   $\langle(\delta\phi)^2\rangle
\sim T_r^2/12 \sim 3\times 10^{-5} \sqrt \lambda
M_{\rm P}^2$. Meanwhile particles created during preheating have much smaller
energy  $\sim  10^{-1} \sqrt \lambda
M_{\rm P}$. Therefore if the same energy density ${\lambda\over 4}10^{-4}
 M_{\rm P}^4$ is instantaneously transferred to low-energy particles created
during preheating, their number, and, correspondingly, the amplitude of
fluctuations, will be much greater,    $\langle(\delta\phi)^2\rangle\sim C^2
M_{\rm P}^2$,  where $C^2 \sim 10^{-2} - 10^{-3}$ \cite{KLS}.   Thermal
fluctuations would lead to symmetry restoration in our
model only
for $\phi_0 {\
\lower-1.2pt\vbox{\hbox{\rlap{$<$}\lower5pt\vbox{\hbox{$\sim$}}}}\
}T_r \sim
10^{-2} \lambda^{1/4} M_{\rm P}  \sim 10^{14}$ GeV  for the realistic value
$\lambda \sim 10^{-13}$ \cite{MyBook} .   Meanwhile, according to eq.
(\ref{p4}),   the
nonthermalized
fluctuations $\langle(\delta\phi)^2\rangle\sim M_{\rm P}^2$  may lead to
symmetry
restoration  even if
the symmetry
breaking parameter $\phi_0$  is  as large as $10^{-1} M_{\rm P}$. Thus,  the
nonthermal
symmetry restoration may occur  even in those theories where the symmetry
restoration due to  high temperature effects would be
impossible.

In reality thermalization   takes a  very  long time, which is inversely
proportional to coupling constants. This dilutes the energy density, and the
reheating temperature becomes many orders of magnitude smaller than $10^{14}$
GeV  \cite{MyBook}. Therefore post-inflationary thermal effects typically
cannot restore symmetry on the GUT scale. Preheating is not instantaneous as
well, and therefore the fluctuations produced at that stage are smaller than
$C^2 M_{\rm P}^2$, but only logarithmically: $\langle(\delta\phi)^2\rangle\sim
C^2 M_{\rm P}^2\ln^{-2}{1\over \lambda}$ \cite{KLS}. For $\lambda \sim
10^{-13}$ this means than nonthermal perturbations produced at reheating may
restore symmetry on the scale up to $\phi_0 \sim 10^{16}$ GeV.

Later   $\langle(\delta\phi)^2\rangle$ decreases as $a^{-2}(t)$  because of the
expansion of the universe. This leads
to  the  phase transition  with symmetry breaking at the moment
$t=t_c\sim \sqrt{\lambda}M_{\rm P}m^{-2}$ when
 $m_{\phi, eff}=0,~
\langle (\delta \phi)^2\rangle=\phi_0^2/3,~E_{\phi}\sim m$. Note
that the homogeneous component $\phi (t)$ at this moment is significantly
less than $\sqrt{\langle (\delta\phi)^2\rangle}$ due to its decay
in the regime of the narrow parametric resonance after
 preheating  \cite{KLS}:
$\overline {\phi^2} \propto t^{-7/6} \propto t^{-1/6} \langle (\delta \phi)^2
\rangle$;  bar means averaging over oscillations.

The mechanism of symmetry restoration  described above is very general; in
particular, it explains a surprising behavior of  oscillations of the scalar
field found  numerically  in the
$O(N)$-symmetric model of ref. \cite{Boyan}. Thus in the interval between
preheating and thermalization the universe could experience a
series of phase transitions which we did not anticipate before. For example,
cosmic strings and textures, which could be an additional source  for the
formation of the large scale
structure of the universe,  should have $\phi_0 \sim 10^{16}$ GeV
\cite{VilBook}. It is   hard to produce them by thermal phase transitions
after inflation
 \cite{KL}.    Meanwhile, as we see now, fluctuations produced at  preheating
may be quite sufficient to restore the symmetry. Then the topological defects
are produced in the standard way  when the symmetry breaks down again. In other
words, production of
superheavy topological defects  can be easily compatible with inflation.

On the other hand, the topological defect production can be quite dangerous.
For example, the model
(\ref{p1})
of a
one-component real scalar field $\phi$ has a discrete symmetry $\phi
\to -
\phi$.
As a result, after the phase transition induced by fluctuations
$\langle(\delta\phi)^2\rangle$ the universe may become filled with
domain
walls
separating phases $\phi = +\phi_0$ and $\phi = -\phi_0$. This
is expected to lead to a cosmological disaster.

This question requires a more detailed analysis. Even though the
point $\phi = 0$ after preheating becomes a minimum of the effective
potential,
the field $\phi$ continues oscillating around
this minimum. Therefore, at the moment $t_c$ it may
happen to  be either
to the right of the maximum of $V(\phi)$ or to the left of it everywhere in
the universe.
In this case the symmetry breaking will occur
in one preferable direction, and no
domain walls will be produced. A similar mechanism may suppress
production of other topological defects.

However, this would be correct only if the magnitude of fluctuations
$(\delta\phi)^2$ were smaller than the average amplitude of the oscillations
$\overline {\phi^2}$.  In our case fluctuations
$(\delta\phi)^2$   are greater than $\overline {\phi^2}$ \cite{KLS}, and they
can have
considerable local deviations from their average value
$\langle(\delta\phi)^2\rangle$. Investigation of this question
shows that  in the theory (\ref{p1})  with $\phi_0 \ll 10^{16}$ GeV
fluctuations destroy the coherent distribution of the oscillating
field $\phi$
and divide the universe into equal number of domains with $\phi = \pm \phi_0$,
which leads to the domain wall problem. This means that in
consistent inflationary models of the type of (\ref{p1}) one  should  have
either $\phi_0 = 0$ (no symmetry breaking), or   $\phi_0 {\
\lower-1.2pt\vbox{\hbox{\rlap{$>$}\lower5pt\vbox{\hbox{$\sim$}}}}\ } 10^{16}$
GeV.

 Now we will consider models where the symmetry breaking occurs for
fields other than the inflaton field $\phi$.
The simplest model  has the following  potential   \cite{KL,Hybrid}:
\begin{equation}\label{p7}
V(\phi,\chi) =   {\lambda\over 4} \phi^4
+ {\alpha\over 4}\Bigl(\chi^2  - {M^2\over \alpha}\Bigr)^2 + {
1\over2} g^2 \phi^2 \chi^2 \ .
\end{equation}
 We will assume here that   $\lambda \ll \alpha, g^2$, so
that
at large $\phi$ the curvature of the potential in the
$\chi$-direction is much
greater than in the $\phi$-direction. In this case at large $\phi$
the field
$\chi$ rapidly rolls toward  $\chi = 0$.
 An interesting feature of such models is the symmetry
restoration for the field $\chi$ for $\phi > \phi_c = M/g$, and
symmetry
breaking when the inflaton field $\phi$ becomes smaller than
$\phi_c$. As was
emphasized in \cite{KL}, such phase transitions may lead to formation
of
topological defects without any need for high-temperature effects.

Now we would like to point out some other specific features of such
models. If
the phase transition discussed above happens during inflation
\cite{KL}  (i.e.
if $\phi_c >
M_p$ in our model), then no new phase transitions occur in this model
after
reheating.
However, for $\phi_c \ll M_p$ the situation is much more complicated.
First of
all, in this case the field $\phi$ oscillates with the initial
amplitude $\sim M_p$ (if $M^4 < \alpha \lambda M_p^4$). This means
that each time when the absolute value of the
field becomes smaller than $\phi_c$, the phase transition with
symmetry
breaking occurs and topological defects are produced.
 Then the
absolute value
of the oscillating field $\phi$
again becomes greater than $\phi_c$, and symmetry restores again.
However, this regime does not continue for
a too long time. Within a few dozen  oscillations, quantum fluctuations of the
field
$\chi$ will be generated with the dispersion
$\langle(\delta\chi)^2\rangle \sim C^2 g^{-1}\sqrt\lambda M^2_{\rm P}
\ln^{-2}{1\over g^2}$ \cite{KLS}.  For $M^2<C^2 g^{-1}\sqrt{\lambda}\alpha
M_p^2\ln^{-2}{1\over g^2}$,
these fluctuations will keep the symmetry restored. Note that this effect may
be even stronger if instead of the term  ${\lambda\over 4} \phi^4$ we would
consider ${m^2\over 2} \phi^2$, since in that case the resonance is more broad
\cite{KLS}. The symmetry breaking
finally completes  when $\langle(\delta\chi)^2\rangle$   becomes
small enough.

One may imagine even more complicated scenario when oscillations
of the
scalar field $\phi$ create large fluctuations of the field $\chi$,
which in
their turn interact with the scalar fields $\Phi$ breaking symmetry
in GUTs.
Then we would have phase transitions in GUTs induced by the
fluctuations of the
field $\chi$.  This means that no longer can the absence of primordial
monopoles be
considered as
an automatic consequence of inflation. To avoid the monopole production one
should  use the theories where   quantum fluctuations produced during
preheating are small or decoupled from the GUT sector. This condition imposes
additional constraints on realistic inflationary models. On the other hand,
preheating may remove some previously existing constraints on inflationary
theory.  For example, in the models of GUT baryogenesis it was assumed that
the GUT symmetry was restored by high temperature effects, since otherwise the
density of X, Y, and superheavy Higgs bosons would be very small.  This
condition is hardly compatible with inflation. It was also required that the
products of decay of  these particles should stay out of thermal equilibrium,
which is a very restrictive condition. In our case the superheavy particles
responsible for
baryogenesis can be abundantly produced by parametric resonance, and the
products of their decay will not be in a state of thermal equilibrium until the
end of reheating.

Now let us  return to   the theory (\ref{p1}) including the field
$\chi$ for
$g^2 \gg \lambda$. In this case the main fraction of the
potential energy density $\sim \lambda M^4_{\rm P}$ of the field $\phi$
predominantly
transfers to the energy of fluctuations of the field $\chi$
due to the explosive $\chi$-particles creation in the broad
parametric resonance.
The dispersion of fluctuations after preheating is
$\langle(\delta\chi)^2\rangle \sim C^2 g^{-1}\sqrt\lambda M^2_{\rm P}
\ln^{-2}{1\over g^2}$. These
fluctuations lead to the symmetry restoration
 in the theory (\ref{p1}) with
$\phi_0 \ll C \Bigl({g^2\over \lambda}\Bigr)^{1/4}  M_p\ln^{-1}{1\over g^2}$,
which may be much greater than $10^{16}$ GeV for $g^2 \gg \lambda$.

Later the process of decay of the field $\phi$ in this model continues, but
less efficiently, and because of the expansion of the universe the fluctuations
$\langle(\delta\chi)^2\rangle$ decrease approximately as
$g^{-1}\sqrt\lambda M^2_{\rm P}\,
 ({a_i\over a(t)})^2$, whereas their energy density $\rho$  decreases as
the energy
density of ultrarelativistic matter,  $\rho(t)  \sim \lambda M^4_{\rm P}\,
({a_i\over a(t)})^{4}$, where $a_i$ is the scale factor at the end of
inflation. This energy density becomes equal to the vacuum energy
density
${m^4\over 4\lambda}$ at $a_0 \sim a_i \, \sqrt\lambda M_p/m,~
t\sim \sqrt{\lambda}M_pm^{-2}$. Since
that time
and until the time of the phase transition with symmetry breaking the
vacuum
energy dominates, and the universe enters secondary stage of  inflation.

The phase transition with spontaneous symmetry breaking occurs when
$m_{\phi ,eff}=0,~\langle(\delta\chi)^2\rangle = g^{-2} m^2$.
This happens at $a_c = a_i\,
\lambda^{1/4}g^{1/2} M_p/m$. Thus, during this additional
period of ``nonthermal'' inflation
the universe expands ${a_c\over  a_0} \sim \sqrt g\,
\sqrt{\phi_0/m}=(g^2/\lambda)^{1/4}$ times.
This is   greater  than expansion during ``thermal'' inflation
(\ref{p5a}) by the
factor $O(g^{-1/2})$, and in our case inflation
occurs even if $g^4 \ll \lambda$.

In this example we considered the second stage of inflation driven by
the
inflaton field $\phi$. However, the same effect can occur in
theories where
other scalar fields are coupled to the field $\chi$. For example, in
the
theories of the type of  (\ref{p5b}) fluctuations
$\langle(\delta\chi)^2\rangle$
produced at the first stage of reheating by the oscillating inflaton
field
$\phi$ lead to a secondary ``nonthermal'' inflation driven by the potential
energy
of the
``flaton'' field $\Phi$. During this stage the universe expands $\sim
\sqrt g\,
 \sqrt{\Phi_0/m}$ times. To have a long enough inflation one may
consider,
e.g., supersymmetric theories with   $m\sim 10^2$ GeV and $\Phi_0
\sim 10^{12}$
\cite{LS}. This gives a relatively long stage of inflation with
${a_c\over
a_0} \sim \sqrt g \ 10^5$, which may be enough to solve the Polonyi
field
problem if the constant $g$ is not too small.

If the coupling constant $g$ is sufficiently
large,
fluctuations of the field $\chi$ will thermalize during this
inflationary
stage. Then the end of this stage will be determined by the standard
theory of
high temperature phase transition, and the degree of expansion during
this
stage will be given by $10^{-1} g\,  \sqrt{\Phi_0/m}$, see eq. (\ref{p5a}). It
is
important,
however, that the inflationary stage may begin even if the field $\chi$
has not been thermalized at that time.

 The stage of inflation described above occurs  in the theory with a
potential which
is not particularly flat  near the origin. But what   happens in the
models
which have flat potentials, like the original new inflation model in
the
Coleman-Weinberg theory \cite{new}?
One of the main problems of inflation in such models was to
understand why
should the scalar field $\phi$ jump onto the top of its effective
potential,
since this field in realistic inflationary models is extremely weakly
interacting and, therefore, it could not be in the state of thermal
equilibrium in
the very early universe. Thus, it is much more natural for inflation
in the
Coleman-Weinberg theory to begin at very large $\phi$, as in the
simplest
version of chaotic inflation in the theory $\lambda \phi^4$.   However,
during the first few oscillations of the scalar
field $\phi$ at the end of inflation
in this model, it produces large nonthermal perturbations of vector
fields
$\langle (\delta A_\mu)^2 \rangle \sim C^2 g^{-1}\sqrt\lambda M^2_{\rm P}
\ln^{-2}{1\over g^2}$. This
leads to
symmetry restoration  and initiates the second stage of inflation
beginning at $\phi = 0$.
It suggests that in many models inflation most naturally begins at
large
$\phi$ as in the simplest version of the chaotic inflation scenario
\cite{Chaotic}. But then, after the stage of preheating, the
second
stage of inflation may begin like in the new inflationary scenario.
In other
words, the new theory of reheating after chaotic inflation may
rejuvenate the
new inflation scenario!

 The main conclusion of this paper is the following. In
addition to the standard high temperature phase transition, there exists
a  new class of phase transitions which
 may occur
 at the intermediate
stage between the end of inflation and the
establishing of thermal
equilibrium. These phase transitions
 take place  even if the scale of symmetry breaking is very large and the
reheating temperature is very small.
Therefore,
phase transitions of the new
type may
have dramatic consequences for inflationary models and the theory
 of physical processes in
the very
early universe.

L.K. and A.L. are very grateful to the organizers of the Workshop on Inflation
at the
 Aspen Center of Physics.
A.L.  was supported in part
by NSF
grant PHY-8612280. A.S.  was supported in part  by the Russian
Foundation
for Basic Research, grant 93-02-3631, and by the Russian Research
Project ``Cosmomicrophysics''.

\end{document}